\documentclass[10pt,twocolumn,epsfig]{article}
\usepackage[dvips]{graphicx}
\def\be{\begin{equation}}
\def\ee{\end{equation}}
\def\no{\nonumber}

\begin{document}

\begin{center}
{\large Apsidal Motion Determination of the Eclipsing Binary Star
V459 Cassiopeiae}
 {\small

{\bf A. DARIUSH}\\
{\it  Institute for Advanced Studies in Basic Sciences (IASBS), Zanjan, Iran.\\E-mail: dariush@iucaa.ernet.in}\\ \ \\
}
\end{center}

\subsection*{\centering Abstract}
\vspace*{-3mm}{\small Photometric and spectroscopic observations
and analysis of the eccentric eclipsing binary V459 Cassiopeiae
($e=0.0244$) were performed by Lacy et al. (2004). Observations of
minimum light show the presence of apsidal motion.  In order to
find the observed rate of apsidal motion, I followed the procedure
described by Guinan \& Maloney (1985). A new observed rate of
apsidal motion of 15$^\circ$.2/100 yr with a period of 2400 yr is
computed which is not in agreement with the one reported earlier.
Also the advance of the periastron is calculated theoretically by
taking into account the Newtonian (classical) and
general-relativistic effects according to the physical and orbital
parameters of the system. The theoretical value of
2$^\circ$.64/100 yr is obtained which is 5.75 times smaller than
the observed rate of the apsidal motion.}\\ \ \\
{\bf Key words:} stars: binaries: eclipsing: V459 Cas -- stars:
apsidal motion


\section{Introduction}
V459 Cassiopeiae (BV5, GSC 04030-01001; $V_{max}=+10.33$; P=8.45
days) is a double-lined eclipsing binary where its variablity  was
discovered by Strohmeier (1955). A measurement on the eccentricity
of the system was done by Busch (1976) after Meinunger \& Wenzel
(1967) recognized the eccentricity of its orbit. Through a series
of spectroscopic and photometric observations (in UBV and $ubvy$)
carried out from 1984 to 2004, the absolute dimensions of the
physical and orbital parameters of V459 Cas was obtained after
precise, simultaneous analysis of the light and velocity curves of
the system (Lacy et al., 2004; hereafter L04). According to L04,
the corresponding radii and masses of the components are
$R_1=2.009\pm0.013R_{\odot}$, $R_2=1.965\pm0.013R_{\odot}$ and
$M_1=2.02\pm0.03M_{\odot}$, $M_2=1.96\pm0.03M_{\odot}$
respectively with an eccentricity of $e$=0.0244. But the rate of
apsidal motion, $\dot{\omega}$, or the period of apsidal
revolution, $U$, is not well determined  which is an important
parameter for probing the stellar structure of the eclipsing
binary components. I investigate the observed rate of apsidal
motion in Sec. 2 by following the method of Guinan \& Maloney
 (1985)(hereafter GM85) to measure the observed rate of apsidal motion,
$\dot{\omega}^{obs}$. The comparison between the theoretical and
the observational values of $\dot{\omega}$ is presented in Sec. 3.
The final results and conclusions are given in Sec. 4.


\section{Observed rate of apsidal motion}
Due to the deep, narrow eclipses of V459 Cas the rate of apsidal
motion can be determined by analysis of primary and secondary
eclipse timings. In their paper, GM85 described the procedure that
must be followed to determine the apsidal motion rate from the
change in the displacement of the
 secondary minimum from the half point (0.5 phase) according to
\begin{equation}
\label{D} D=(t_{2}-t_{1})-0.5{\times}Period,
\end{equation}
where $t_2$ and $t_1$ are times of secondary and primary minima,
respectively and $D$ is related to the longitude of periastron
$\omega$ by the formula given by Sterne (1939,a,b)
\begin{equation}
\label{DW}
 D=\frac{P}{\pi}\left[\tan^{-1}\left(
\frac{e\cos{\omega}}{{(1-e^2)}^{1/2}}\right) +
\frac{e\cos{\omega}} {1-{e^2}{\sin^2{\omega}}}
{(1-e^2)}^{1/2}\right].
\end{equation}

\begin{figure}
\label{oc}
\begin{center}
\includegraphics[scale=.7]{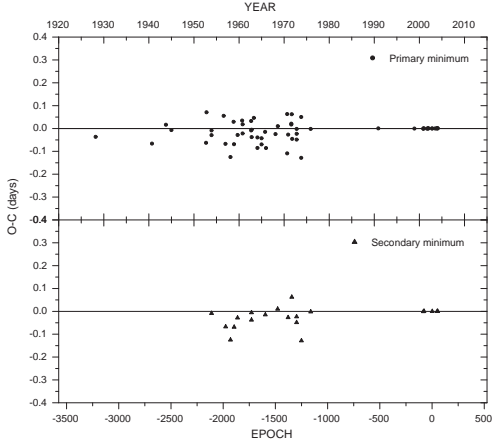}
\caption{Differences between the observed and the computed (O-C)
times of primary and secondary minima, computed  from Eq.
(\ref{ef}), versus Julian Day number and year. The observed times
of secondary minima are listed in Table 1.}
\end{center}
\end{figure}

\begin{figure}
\label{wjd}
\begin{center}
\includegraphics[scale=.85]{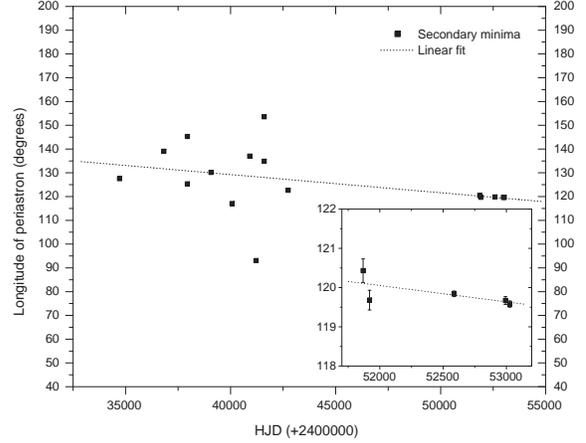}
\caption{ Plot of the longitude of periastron ($\omega$) versus
Julian day number according to the displacement of observed
secondary minima listed in Table 1 from a half period point ($D$).
The inset shows only the last five minima of Table 1 for which the
calculated slope of the diagram is based only on these five
secondary times of minima. A least square fit to the observations,
yielding a slope $\dot{\omega}^{obs}=15.18$deg/100 yr.}
\end{center}
\end{figure}

\begin{figure}
\label{dww}
\begin{center}
\includegraphics[scale=.7]{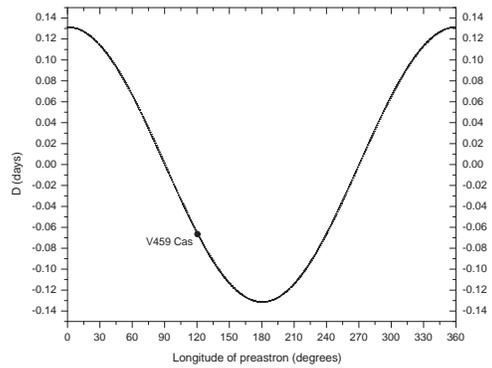}
\caption{$D$ as a function of $\omega$ using Eq. (\ref{DW}).}
\end{center}
\end{figure}

\begin{figure}
\label{orbit}
\begin{center}
\includegraphics[scale=.3]{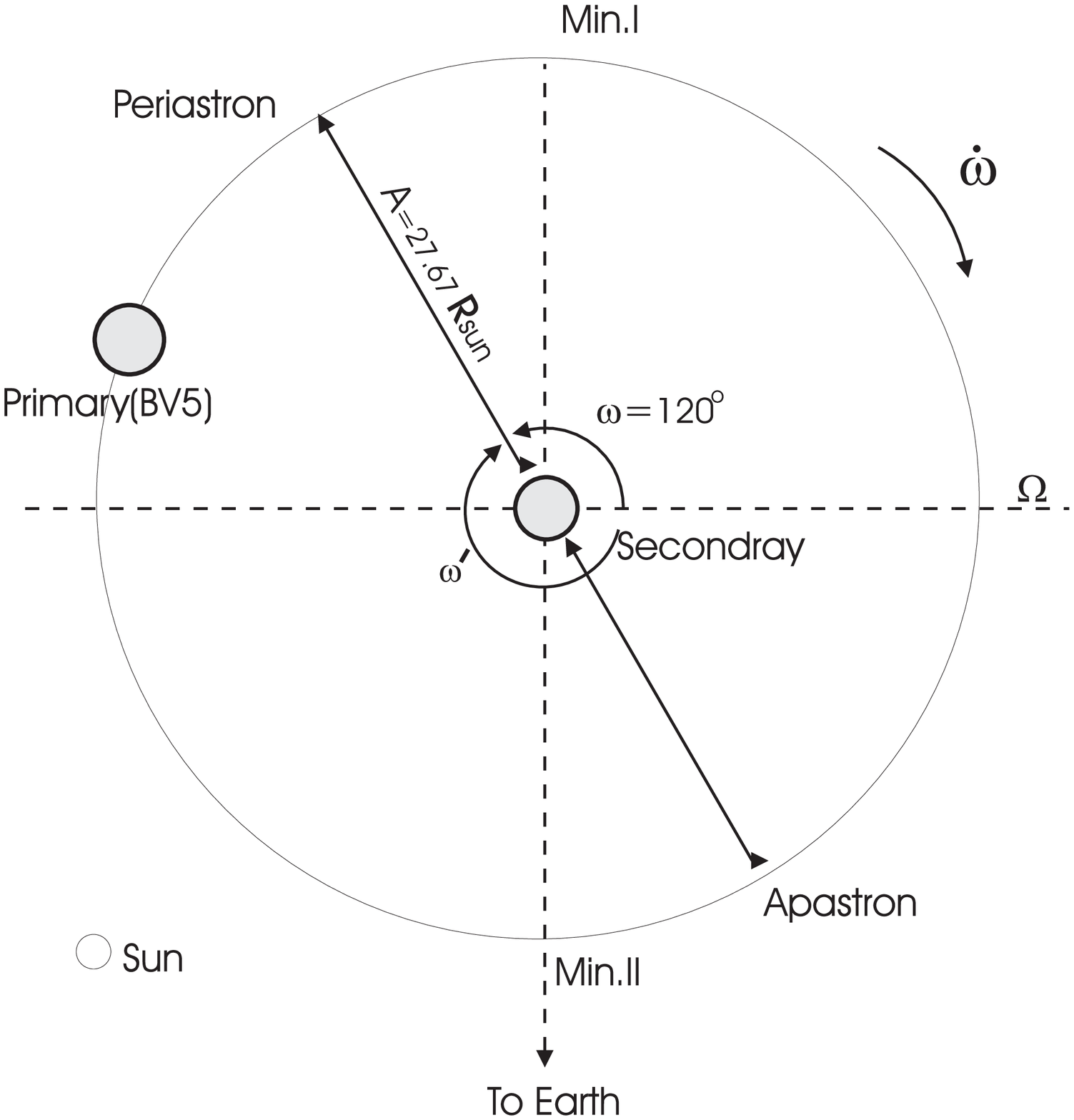}
\caption{Relative orbit of V459 Cas, drawn to scale, from the
quantities given by Lacy et al.(2004). }
\end{center}
\end{figure}

In Eq.  (\ref{DW}), $e$ is the orbital eccentricity and $P$ is the
orbital period in days. Fig. 1, presents the (O-C) diagram from
all the observed times of primary and secondary minima. To plot
the (O-C) diagram for the primary minima (upper panel of Fig. 1),
I used the times of minima listed by L04 as well as two other
minima reported by Lacy et al. (1998) and H$\ddot{u}$bscher
(2005). The residuals are calculated according to the ephemeris
given by L04
\begin{equation}
\label{ef} Min.I=HJD \hspace{2mm}2452565.67170+8.45825381d \times
E.
\end{equation}
For all the secondary minima in Table 1, $D$ and from that the
$\omega$ are calculated using Eq. (\ref{DW}). As you can see from
Table 1, the slow increase (or decrease) in $D$ (or $\omega$) is
due to the advance of the line of apsides of the orbit and by
computing the slope of a line fitted to  the secondary minima on
${\omega}$ versus Julian Day number diagram, I determined a new
rate of apsidal motion $\dot{\omega}^{obs}=15^\circ.18 \pm 6.75$
(see Fig. 2).

It must be noted that among the observed secondary minima listed
in Table 1, the error values of the last five minima are at least
100 times smaller than the other minima. Therefore, I use only
these five minima to measure the $\dot{\omega}^{obs}$. Also in
some cases the error in secondary minima are so high and Eq.
(\ref{DW}) doesn't give any $\omega$ correspond to the $D$,
calculated from Eq. (\ref{D}) since as it is obvious from Fig. 3,
for any set of values of $e$ and $P$ in Eq. (\ref{DW}), $D$ is
always limited between its maximum and minimum values (e.g.
$D_{min}<D<D_{max}$). Thus, I can not find any $\omega$ correspond
to $D>D_{max}$ or $D<D_{min}$. Finally Fig. 4 shows the relative
orbit of V459 Cas, drawn to scale. Actually what can be measured
from Eq. (\ref{DW}) is $\omega$ and according to Fig. 4,
$\dot{\omega}^{obs}<0$ but L04 have considered $\omega'$ as the
angle of line of apside. One can reveal from Fig. 4 that
$\omega=2\pi-\omega'$ and
$|\dot{\omega}^{obs}|=|\dot{\omega'}^{obs}|$.

\section{Comparison with theory}
Basically the theoretical rate of apsidal motion is due to the
contribution of two terms; a classical term as well as the
general-relativistic term. Then the total rate of apsdial motion
is equal to
\begin{equation}
\label{total}
\dot{\omega}_{tot}^{theo}=\dot{\omega}_{CL}^{theo}+\dot{\omega}_{GR}^{theo},
\end{equation}

where $\dot{\omega}_{CL}^{theo}$ denotes the classical or
Newtonian term and $\dot{\omega}_{GR}^{theo}$ is the relativistic
contribution. I am going now to calculate each term in equation
(\ref{total}) with the assumption that the rotation equators of
both stars are coincident with their orbital plane. In classical
mechanics, the advance of the line of apsides arises due to the
lack of spherical symmetry in the shape of the stars. Tidal and
rotational distortion are two mechanisms lead to this asymmetry
which depend on the fractional radii, the internal mass
distribution, and the axial rotation of the stars.
$\dot{\omega}_{CL}^{theo}$ can be determined using the expressions
given by Cowling (1938), Sterne (1939), and Kopal (1959)

\begin{eqnarray}
\no\dot{\omega}_{CL}^{theo}(\frac{deg}{yr})&=&365.25\big(\frac{360}{P}\big)
\no\big\{k_{2,1}{r_1}^5[15f_2(e)(\frac{M_2}{M_1})\\
\no&+&(\frac{{\tilde{\omega}}_{r,1}}{{\tilde{\omega}}_{k}})^2(\frac{1+M_2/M_1}{{(1-e^2)}^2})]\\
\no&+&k_{2,2}{r_2}^5[15f_2(e)(\frac{M_1}{M_2})\\
&+&(\frac{{\tilde{\omega}}_{r,2}}{{\tilde{\omega}}_{k}})^2(\frac{1+M_1/M_2}{{(1-e^2)}^2})]\big\},
\label{cl}
\end{eqnarray}

where $f_2(e)=(1+\frac{3}{2}e^2)(1-e^2)^{-5}$, $P$ is the orbital
period in days, $M_1$ and $M_2$ are the masses of stars in terms
of $M_{\odot}$, $k_{2,1}$ and $k_{2,2}$ are known as the apsidal
motion constants of the component 1 and 2 respectively, $r_1$ and
$r_2$ are the fractional radii of the stars ($r=\frac{R}{A}; A$=
semimajor axis of the orbit), $\tilde{\omega}_{r,1}$ and
$\tilde{\omega}_{r,2}$ are the star's angular rotation speeds, and
$\tilde{\omega}_k$ is the mean angular Keplerian velocity equal to
$\frac{2\pi}{P}$. In series expansion $f_2(e)$, I did not keep the
terms involving $e^4$ and higher orders since they have not any
significant contribution to the result of summation in $f_2(e)$.
L04 indicated that the components of V459 Cas are main-sequence
stars with an age of about 525 Myr. Therefore in order to
calculate $k_2$ which is correspond to the second harmonic of the
mutual tidal distortion, I used Table 1 of Jeffery (1984) where he
has computed $k_2$ from modern stellar-interior models of evolving
main sequence stars.
 From the estimated values for chemical composition ($X=0.712, Z=0.012$), I adopt
 a nearly
 equal value of $k_{2,1}\simeq k_{2,2}$=0.0042 for both components.
Using the profiles of absorption lines in the star's spectrum, it
is possible to estimate the angular rotation velocities
$\tilde{\omega}_r$ of the stars from their projected rotational
velocity, $V_r\sin i$. L04 have estimated $V_r\sin i$ from the
absorption lines of
 Mg$_{II}\lambda$4481 in the spectrogram of V459 Cas. From these
 measurements, $V_r\sin i$=54 $kms^{-1}$ and 43 $kms^{-1}$ for the first and the second
components respectively. Therefore
$\tilde{\omega}_{r,1}/\tilde{\omega}_k$=4.49 and
$\tilde{\omega}_{r,2}/\tilde{\omega}_k$=3.66 are obtained from
$\tilde{\omega}_{r,i}=V_{rot}/{R_i}$. Substituting all of the
parameters and the values from L04 into equation
(\ref{cl}), I find\\

$\dot{\omega}_{CL}^{theo}=1^\circ.22\pm0.12$/100 yr.\\ \ \\
The expression for the advance of the line of apsides due to
general relativity has a simple form and according to Levi-Cevita
(1937) and Kopal (1959) is equal to

\begin{equation}
\label{gr}
\dot{\omega}_{GR}^{theo}(\frac{deg}{yr})=9.2872\times10^{-3}\frac{{(M_1+M_2)}^{2/3}}{{(P/2\pi)}^{5/3}(1-e^2)},
\end{equation}
where $M_1$ and $M_2$ are in $M_{\odot}$ and $P$ is in days.
Putting again the value of the parameters into equation (\ref{gr})
yields\\

$\dot{\omega}_{GR}^{theo}=1^\circ.42\pm0.01$/100 yr.\\ \ \\
Thus the combined classical and general relativistic rate of
apsidal motion is equal to\\

$\dot{\omega}_{CL+GR}^{theo}=2^\circ.64\pm0.12$/100 yr,\\ \ \\
which is calculate according to equation (\ref{total}). To
calculate the standard deviations, I used the errors assigned to
each parameter given in the references. The above value indicates
that in comparison to $\dot{\omega}^{obs}$,
$\dot{\omega}_{CL+GR}^{theo}$ is about 5.75 times smaller with a
discrepancy of $12^\circ.54$deg/100 yr.


\section{{Results and discussion}}

V459 Cas has been observed and analyzed carefully by L04. In this
paper I followed the method of GM85 to measure a value for the
observed rate of apsidal motion. Table 2 presents the results of
independent determinations of the rate of apsidal motion together
with  the corresponding period of apsidal revolution $U$. The
relation between the apsidal motion period $U$ and
$\dot{\omega}^{obs}$ has a simple form
\begin{equation}
U=\frac{360P}{\dot{\omega}_{obs}},
\end{equation}
where $\dot{\omega}_{obs}$ is expressed in degrees per cycle and
$P$ is the anomalistic period expressed in days. Also the observed
rate of apsidal motion is 5.75 times greater from the expected
theoretical value. It is not the first time that such a
discrepancy is observed. For example in the case of V1143 Cygni
the observed rate of apsidal motion is about one-third of the
value estimated from the theory but according to the new studies
it seems that this deviation is decreasing with the expansion of
observational time (Dariush et al., 2004). On the other hand in
the case of highly eccentric eclipsing binary, DI Herculis, the
observed rate of apsidal motion is one-seventh of the estimated
theoretical value. Tough a series of investigations have been made
during the past decades to find a clear explanation for this high
discrepancy (GM85; Claret, 1998; and Claret \& Willems, 2002) but
the case of DI Herculis is still unsolved.
 The method of GM85 is expected to gives the
value of $\dot{\omega}^{obs}$ exactly but to determine a more
accurate value for $\dot{\omega}^{obs}$ we need more accurate
timings of secondary minima since up to now, our observational
baseline covers only a very small fraction of the period of
apsidal revolution $U$. Therefore it seems that the reason for
such a large discrepancy between the theory and the observation is
mainly due to the small number of observational data points.


\subsection*{\centering Acknowledgements}
This research has made use of {\bf NASA}'s Astrophysics Data
System Abstract Service and {\bf SIMBAD} data base operated at CDS
(Strasbourg, France).

The author also gratefully acknowledge the valuable help of Mr.
Hadi Rahmani with some of the calculations and the financial
support of the Institute for Advanced Studies in Basic Sciences
(IASBS).


\begin{table}
\caption{The photoelectric times of secondary minima for V459 Cas,
together with  the  values of $D$ and $\omega$, computed using
Eqs. (\ref{D}) and (\ref{DW}) according to the
light element given in Eq. (\ref{ef}).} 
\label{table:1} 
\centering 
\begin{tabular}{l r r l l} 
\hline\hline 

H.JD.        & error  & Epoch &    D    & $\omega$(deg) \\
(2400000.+)  &           &       &         &       \\ \hline
34714.447   &   0.05    &   -2110.5   &   -0.08    &   127 \\
35856.253   &   0.05    &   -1975.5   &   -0.13    &   --  \\
36245.275   &   0.05    &   -1929.5   &   -0.19    &   --  \\
36541.37    &   0.05    &   -1894.5   &   -0.13    &   --  \\
36820.533   &   0.05    &   -1861.5   &   -0.09    &   139 \\
37945.472   &   0.05    &   -1728.5   &   -0.10    &   145 \\
37945.504   &   0.05    &   -1728.5   &   -0.07    &   125 \\
39070.443   &   0.05    &   -1595.5   &   -0.08    &   130 \\
40068.542   &   0.05    &   -1477.5   &   -0.05    &   117 \\
40914.331   &   0.05    &   -1377.5   &   -0.09    &   136 \\
41210.459   &   0.05    &   -1342.5   &   -0.01    &   93  \\
41599.428   &   0.05    &   -1296.5   &   -0.11    &   153 \\
41599.453   &   0.05    &   -1296.5   &   -0.09    &   134 \\
41988.428   &   0.05    &   -1250.5   &   -0.19    &   --  \\
42741.339   &   0.05    &   -1161.5   &   -0.07    &   122 \\
51867.7992  &   5E-4    &   -82.5 &   -0.0665    &   120.4  \\
51918.5502  &   4E-4    &   -76.5 &   -0.0650    &   119.6  \\
52586.75193 &   1.3E-4  &   2.5   &   -0.06540    &   119.84  \\
52992.74846 &   1.9E-4  &   50.5  &   -0.06506    &   119.67  \\
53026.58164 &   1.4E-4  &   54.5  &   -0.06489    &   119.58  \\
\hline
\end{tabular}
\end{table}

\begin{table}
\caption{Determined rate of apsidal motion for V459 Cas.} 
\label{table:2} 
\centering 
\begin{tabular}{l l l l } 
\hline\hline 
$\dot{\omega}^{obs}$ (deg/100 yr) & U$^{yr}$ & Source \\ \hline
$6.04 \pm 4.74$ & $6100$  & L04 \\
$15.18 \pm 6.75$ & $2400$  & Present study \\ \hline
\end{tabular}
\end{table}

\end{document}